\documentclass[%
 reprint,
showpacs,
 amsmath,amssymb,
 aps,prb,
 longbibliography,
 lengthcheck,%
]{revtex4-1}

\usepackage{graphicx}
\usepackage{dcolumn}
\usepackage{bm}
\usepackage{hyperref}

\begin{document}

\preprint{APS/123-QED}

\title{Optical response of small closed-shell sodium clusters}

\author{George Pal}
\affiliation{Physikalisch-Technische Bundesanstalt (PTB), Bundesallee 100, 38116 Braunschweig, Germany}

\author{Georgios Lefkidis}%
 \email{lefkidis@physik.uni-kl.de}
\author{Hans~Christian~Schneider}
\author{Wolfgang~H\"{u}bner}
\affiliation{%
Physics Department and Research Center OPTIMAS, University of Kaiserslautern, P.O.Box 3049, 67653 Kaiserslautern, Germany
}%

\date{\today}

\begin{abstract}
  Absorption spectra of closed-shell Na$_2$, Na$_3^+$, Na$_4$,
  Na$_5^+$, Na$_6$, Na$_7^+$, and Na$_8$ clusters are calculated using
  a recently implemented conserving linear response method. In the
  framework of a quasiparticle approach, we determine electron-hole
  correlations in the presence of an external field. The calculated
  results are in excellent agreement with experimental spectra, and
  some possible cluster geometries that occur in experiments are
  analyzed.  The position and the broadening of the resonances in the
  spectra arise from a consistent treatment of the scattering and
  dephasing contributions in the linear response
  calculation. Comparison between the experimental and the theoretical
  results yields information about the cluster geometry, which is not
  accessible experimentally.
\end{abstract}

\maketitle


\section{Introduction}

The electronic configuration of small metal clusters is
closely connected with their geometrical, chemical and optical
properties~\cite{Knight1}. Different aspects of this interrelation for small metal clusters have been extensively studied experimentally~\cite{Knight2Exp,
Kappes1Exp, Kappes2Exp, Knight3Exp, deHeer, Haberland3Exp,
Haberland1Exp, Haberland2Exp, Haberland4Exp, IssendorffExp,
BalettoExp} and theoretically~\cite{ClemengerShell1985,
BeckJellium1984, EckardtJellium1984, BrackJellium1993, 
ReinhardJellium2000, BortignonExp, Koutecky31988, Koutecky11990, Koutecky21990,
Koutecky41992, Koutecky51996, RubioPRL1996, OgutPRL1999,
RubioJCP2001, UziPRL2001, Broglia2005, NieminenDFT2008,
OnidaNa4PRL1995, Louie1PRB2000, OnidaRMP2002, LopezPRB2008,
Ogut2PRB2009}. For instance, the measured photoabsorption cross
section of mass-selected clusters yields important information about
the interplay between the underlying nuclear configuration
and the electronic structure. However,
the geometry of the clusters in the gas phase is not directly accessible in experiments.
 \emph{Ab initio} theory can help to determine the
structures by choosing, among many
possibilities, the most stable geometrical configuration, for which
the calculated absorption spectrum is in good agreement with the
experimental results.

Early theoretical attempts to understand the nature of electronic
correlation in small alkali-metal clusters are based on the
shell~\cite{ClemengerShell1985} and density functional theory
(DFT)-based jellium~\cite{BeckJellium1984, EckardtJellium1984,
  BrackJellium1993, ReinhardJellium2000} models.  Quantum molecular
approaches based on the Configuration Interaction (CI) procedures are
powerful theoretical tools capable of yielding accurate electronic
excited states and important information about the nature of the
optically allowed transitions between them. The multi-reference single
and double excitation CI has been successfully used to account for the
dominant features of the absorption spectra of small sodium
clusters~\cite{Koutecky31988, Koutecky11990, Koutecky21990,
  Koutecky41992, Koutecky51996}, calculated from the oscillator
strength of the transitions between the ground and the excited
states. Also, time-dependent (TD) DFT methods are widely employed to
compute the photoabsorption cross sections of small metal clusters,
with various levels of sophistication for the exchange-correlation
functional~\cite{RubioPRL1996, OgutPRL1999, RubioJCP2001, UziPRL2001,
  Broglia2005, NieminenDFT2008}.  A different way to compute optical
spectra is to make use of the quasiparticle picture, where a typical
photoabsorption process involves an incoming photon and the transition
of an electron from an occupied to an empty quasiparticle state. The
determination of the cross section can be achieved by a calculation of
a generalized ``four-point`` (i.e., two-particle) electron-hole
correlation function, which obeys a Bethe-Salpeter equation
(BSE)~\cite{OnidaNa4PRL1995, Louie1PRB2000, OnidaRMP2002,
  LopezPRB2008, Ogut2PRB2009}.  In practice, this is usually done in a
two-step calculation: first the DFT or Hartree-Fock (HF)
single-particle states are used as input into a GW-like procedure to
obtain the quasiparticle corrections to the Kohn-Sham or HF
ground-state eigenvalues~\cite{HedinGW, LouieNaGW}. Then a
two-particle BS effective equation with a screened direct and an
unscreened exchange electron-hole interaction is solved. While the
first step properly describes charged excitations (i.e., electron
removal or addition, specific to direct or inverse photoelectron
experiments), the second step accounts for neutral excitations (i.e.,
electron-hole pair creation, specific to photoabsorption
experiments). A comprehensive comparison between the TDDFT and the
GW+BSE approaches can be found in Ref.~\citenum{OnidaRMP2002}. To
facilitate the numerical procedure, the screened Coulomb interaction,
which is obtained in the GW step and which enters the BSE, is most
often taken to be the statically screened Coulomb interaction. This
simplification allows one to cast the effective two-particle BSE in
the form of an energy-independent eigenvalue problem, which yields the
excited states corresponding to neutral excitations of the system. As
in the case of CI, the optical absorption cross section is then
determined by the oscillator strengths of the transitions to the
excited states, and a phenomenological parameter for the resonance
peak broadening is added in order to compare the calculated spectra
with the shapes of the experimental peaks.  This phenomenological time
for the individual transition mimics the combination of thermal
broadening and electron lifetime.

We have developed, in a quasiparticle picture, an alternative method
to determine the photoabsorption spectra of small sodium clusters. We
employ a first-principles based linear response calculation for the
dynamical electron-hole correlation function in the presence of an
external potential. Although we do assume a ''background''
quasiparticle lifetime, the finite width of the absorption peaks
contains electron-hole correlation contributions beyond a mere
convolution of single-particle lifetimes.  A novel aspect of our
approach is that we use a quantum kinetic equation for the
electron-hole correlations driven by an external coherent optical
field to obtain a closed equation for the two-particle BSE without the
need of a four-point correlation function.  Technically, we include
mean-field HF contributions (accounting for direct and exchange bare
Coulomb interaction) together with correlation contributions beyond
mean-field, described by complicated integral kernels. An important
difference from the two-step GW+BSE approaches mentioned above is that
we do not include correlations on the single-particle level and then
solve the BSE with an effective two-particle interaction described in
terms of a simple integral kernel. Instead, we follow the original
prescription of deriving conserving approximations for transport
equations by Baym and Kadanoff. We thus obtain a BSE for the
electron-hole correlation function that is consistent with the
electronic single-particle properties~\cite{BaymKadanoff1961}.  This
type of conserving approach (in the sense of Baym and Kadanoff) avoids
possible problems of double-counting correlation contributions which
may arise in a two-step GW+BSE-like calculation due to the separate
treatment of single- and two-particle correlations. As already shown
by Baym and Kadanoff, such a conserving approach fulfills important
sum-rules by construction. For instance, the particle-number
conservation law for the electron-hole correlation function, which can
be read as a sum rule for the absorption cross section in finite
systems, serves as a check of our numerical calculations.

\section{Theory}\label{SectionTheory}

The BS approach employed to calculate the absorption spectra is
described in detail in Refs.~\citenum{PalEPJB}
and~\citenum{PalAbs1}, and is briefly reviewed in the following.
The central quantity of this approach is the retarded
density-density correlation function
$\chi^{\mathrm{r}}(\mathbf{r},t;\mathbf{r^{\prime} },t^{\prime})$,
which describes the linear response of the system to an external
potential $U(\vec{r},t).$ For finite systems, we use the matrix element
\begin{equation}
\langle
n_{1}n_{2}|\chi^{\mathrm{r}}(t,t^{\prime})|n_{3}n_{4}\rangle=\left.
\frac{\delta\rho_{n_{2}n_{1}}(t)}{\delta U_{n_{3}n_{4}}(t^{\prime}%
)}\right\vert _{U=0}
\label{densitydensitycorrelationfunction}%
\end{equation}
in a basis of HF eigenfunctions $\{\varphi_{n}(\mathbf{r})\}$. Here,
$n$ labels the HF spin orbital, and $\rho_{n_{2}n_{1}}(t)=\langle
c_{n_{2}}^{\dag }(t)c_{n_{1}}(t)\rangle$ with the annihilation
(creation)\ operators $c_{n}$ ($c_{n}^{\dag}$) for an electron in HF
orbital $\varphi_{n}.$ In linear response, the electronic
distribution functions $\rho_{nn}$ are unchanged, and the only nonzero
elements in Eq. \eqref{densitydensitycorrelationfunction} are those
with indices that pair occupied  and unoccupied states, i.e.,
Eq. \eqref{densitydensitycorrelationfunction} is the electron-hole
correlation function.

The photoabsorption cross section is calculated from the Fourier
transform of the imaginary part of $\chi^{\mathrm{r}}$  according to
\begin{equation}\label{crosssection_n}
\sigma(\omega)= \frac{\omega}{\varepsilon_{0}c}  \sum_{n_{1}\dots
n_{4}}\mathrm{Im}\langle n_{2}n_{1}|\chi^{\mathrm{r}}(\omega)|n_{3}%
n_{4}\rangle   \, {\mathbf d}_{ n_{1}n_{2}}\cdot {\mathbf
d}_{n_{3}n_{4}} ,
\end{equation}
where ${\mathbf d}_{ n_{1}n_{2}}=\int\varphi_{n_{1}}^{\ast} \,( e\,
{\mathbf r} ) \, \varphi_{n_{2}} d^{3}r$  are the electric dipole
matrix elements in the molecular orbital representation and $e$ is the
electron charge.

The two-particle BS equation for  $\chi^{\mathrm r}$ is
\begin{align}
\big( & \omega- \epsilon_{n_1}^{\mathrm{HF}}
+\epsilon_{n_2}^{\mathrm{HF}} \big)
\langle n_1 n_2 | \chi^{\mathrm r} (\omega) | n_3 n_4 \rangle + \nonumber \\
& (f_{n_1}-f_{n_2}) \Big[ \delta_{n_1n_3} \delta_{n_2n_4}+
\sum_{n_5n_6} \Big( \langle n_1 n_5|v|n_2n_6 \rangle \nonumber
\\ & -  \langle n_1n_2|v|n_5n_6 \rangle \Big) \langle n_5 n_6 |
\chi^{\mathrm r}(\omega) | n_3 n_4 \rangle \Big]   \nonumber \\
 &+  \sum_{n_5n_6} \langle n_1 n_2 | \Delta (\omega) | n_5 n_6
\rangle \langle n_5 n_6 | \chi^{\mathrm r} (\omega) | n_3 n_4
\rangle = 0
    \label{eqnmotchi_w}
    \end{align}
    with the correlation kernel
\begin{align}\label{finalD}
\langle  n_1  n_2 & |\Delta(\omega)| n_3n_4\rangle = \nonumber\\
  \sum_{n_5n_6} \Big[ & \frac{ f_{n_1} (1-f_{n_5}) f_{n_6} + (1-
f_{n_1} ) f_{n_5} (1-f_{n_6})}{ \omega - \tilde
\epsilon^{*}_{n_6} + \tilde \epsilon_{n_5} - \tilde
\epsilon^{*}_{n_1}+ \tilde \epsilon_{n_4} } \nonumber \\  & \qquad \times \langle n_1n_5 | v |
n_3n_6 \rangle \langle n_6n_2 | v | n_5n_4 \rangle  \nonumber
\\
+   & \frac{  f_{n_2} (1-f_{n_5}) f_{n_6} + (1-
f_{n_2} ) f_{n_5} (1-f_{n_6}) }{ \omega + \tilde \epsilon_{n_6}
- \tilde \epsilon^{*}_{n_5} - \tilde \epsilon^{*}_{n_3}+ \tilde
\epsilon_{n_2} } \nonumber \\
 & \qquad  \times \langle n_1 n_5 | v | n_3 n_6 \rangle \langle n_6
n_2 | v | n_5 n_4 \rangle\nonumber
 \\
- \delta_{n_1n_3}  & \sum_{n_7}  \frac{  f_{n_7}
(1-f_{n_6}) f_{n_5} + (1- f_{n_7} ) f_{n_6} (1-f_{n_5})}{
\omega - \tilde \epsilon^{*}_{n_6} + \tilde \epsilon_{n_5} - \tilde
\epsilon^{*}_{n_3}+ \tilde \epsilon_{n_7} } \nonumber \\
 & \qquad  \times  \langle n_4n_5 | v | n_7n_6 \rangle \langle n_6n_2 | v | n_5n_7 \rangle\nonumber
 \\
-\delta_{n_2n_4}  & \sum_{ n_7}  \frac{ f_{n_7}
(1-f_{n_6}) f_{n_5} + (1- f_{n_7} ) f_{n_6} (1-f_{n_5})}{
\omega + \tilde \epsilon_{n_6} - \tilde \epsilon^{*}_{n_5} - \tilde
\epsilon^{*}_{n_7}+ \tilde \epsilon_{n_4} } \nonumber \\
 & \qquad  \times  \langle n_1n_5 | v | n_7n_6 \rangle
\langle n_6n_3 | v | n_5n_7 \rangle   \Big].
\end{align}
In Eqs.~\eqref{eqnmotchi_w} and~\eqref{finalD} $f_n=\rho_{nn}$ are
the level occupation numbers assumed to be equilibrium Fermi-Dirac
distributions at $T=0$\,K. The bare Coulomb matrix element is defined by
\begin{eqnarray} \label{indexCoulomb}
\langle n_1n_2 | v |n_3n_4 \rangle  & =\int d^3r_1 d^3 r_2  \,
\varphi^*_{n_1}({\mathbf r}_1) \varphi^*_{n_2} ({\mathbf r}_2) \\
 & \qquad \times \,   v(|{\mathbf r}_1 - {\mathbf r}_2|) \, \varphi_{n_3}({\mathbf r}_1)
\varphi_{n_4}({\mathbf r}_2).  \nonumber
\end{eqnarray}
The denominators in the expression for the correlation
kernel~\eqref{finalD} contain the quasiparticle energies including an inverse  lifetime
$\tilde \varepsilon_n= \varepsilon_n+i\gamma_n$. In principle, one
can use for $\tilde \varepsilon_n$ the results of a quasiparticle
calculation such as a GW calculation~\cite{YaroslavPRL, PalEPJB,
YaroslavGW}, but here we employ the HF energies and  a
constant broadening for all states whose magnitude is in agreement with GW calculations. We stress  that
introducing a fixed \emph{quasiparticle} lifetime for the
calculation of $\Delta(\omega)$ does not mean that the broadening of
the resonances in~$\chi^{\mathrm r}$  is given by this value.
Rather, the resonance broadening, which is responsible for the
finite width of the peaks in the absorption spectrum, is due to the
imaginary part  of the full
correlation contribution~$\Delta(\omega)$. The real part of
$\Delta(\omega)$ yields the shift in the resonance energies.
Both effects, renormalization and broadening of the
$\chi^{\mathrm r}(\omega)$ spectrum, are thus related.

Equation~\ref{eqnmotchi_w} for the electron-hole correlation can be
solved at different levels of approximation. Neglecting all
Coulomb interactions and setting
\begin{equation}\label{LindhardLimit}
\langle n_1 n_2 |\Delta(\omega)|
n_3n_4\rangle = i \delta_{n_1n_3} \delta_{n_2n_4} \Gamma
\end{equation}
one obtains for $\chi^{\mathrm r}$ the independent particle
(Lindhard) result
\begin{equation}\label{Lindhard}
\langle n_1 n_2 | \chi_{0}^{\mathrm{r}}(\omega) | n_3 n_4 \rangle =
- \frac{ \delta_{n_1n_3}\delta_{n_2n_4}
(f_{n_1}-f_{n_2})}{\hbar(\omega+i\eta) -\epsilon^{\mathrm{HF}}_{n_1}
+ \epsilon^{\mathrm{HF}}_{n_2}}.
\end{equation}
Including direct Coulomb matrix elements,
exchange elements and correlation contributions in that order, Eq.~\eqref{eqnmotchi_w} resembles a BSE in the ladder
approximation (denoted by BS-L in the following) plus exchange
(BS-LX) and correlation terms (BS-LXC), respectively~\cite{PalEPJB}. Recall that we use loosely here the term BSE for Eq.~\eqref{eqnmotchi_w}, even though
the BSE is usually written for the four-point generalization of $\chi$.

\section{Photoabsorption spectra of small Na clusters}\label{SectionResults}

In this section we discuss photoabsorption spectra of  closed-shell Na$_2$, Na$_3^+$,
Na$_4$, Na$_5^+$, Na$_6$, Na$_7^+$, and Na$_8$ clusters. The numerical results are calculated according to Eq. \eqref{crosssection_n}  and compared
with experiment. We choose clusters with closed-shell configurations  with an even number of electrons, for which  all molecular orbitals
are doubly occupied or empty. Moreover, Na$_2$, Na$_3^+$, and Na$_8$ are
magic-number clusters, for which the number of valence electrons equals the
spherical-shell closing-numbers 2 and 8, respectively~\cite{deHeer}.
The electron-hole correlation function
is determined by solving  Eq.~\eqref{eqnmotchi_w} with the
correlation kernel~\eqref{finalD}.

To obtain the equilibrium ($T=0$\,K) geometrical configuration,
we perform a structural optimization using the HF approximation,
which yields the single-particle ground state energies and wave
functions. For the Na atoms we use the \textsc{lanl2dz} basis set
(double-zeta set with relativistic Los Alamos effective core
potential)~\cite{LANL2DZ}, in which the $3s^1$ valence electron of
each Na atom is represented by contracted Gaussian-like atomic
orbitals  (3s3p/2s2p) while the contribution of the core electrons
is treated using effective core potentials (ECP). To test the
stability of our results with respect to the basis set used in the
calculations, we have also performed calculations with six other
basis sets: \textsc{SHC} (Goddard/Smedley ECP)~\cite{SHC},
\textsc{CEP-4G} (minimal set and Stephens/Basch/Krauss ECP),
\textsc{CEP-31G} (split valence set) and \textsc{CEP-121G}
(triple-split basis set)~\cite{CEP}, \textsc{LanL2MB} (MBS set and
Los-Alamos ECP)~\cite{LANL2DZ}, and \textsc{SDDall} (double-zeta set
and Stuttgart/Dresden ECP)~\cite{SDDAll}.  For larger basis sets such as \textsc{lanl2dz},
\textsc{SHC}, \textsc{CEP-121G} and \textsc{SDDall} the Na-Na
distances vary by less than 1\% at the end of the geometry
optimization, and the peak positions of the absorption spectra
differ by less than 0.1~eV. For smaller basis sets the Na-Na
distances vary at most by 0.3~{\AA} while the peak positions have
deviations of less than  0.3~eV. Overall, a good comparison with experiment
is found for \textsc{lanl2dz}, and in the following we refer only to
results obtained using this basis set. When the absorption spectra
are calculated from the density-density correlation function obtained
in the Lindhard, BS-L and BS-LX approximations, we use a constant broadening
$\Gamma=0.03$~eV in Eq.~\eqref{LindhardLimit}. When correlation terms $\Delta$ are included in
the calculations we use for the quasiparticle inverse lifetime  the value
$\gamma=0.15$~eV in Eq.~\eqref{finalD}.

\begin{figure}[htb!]
\includegraphics[scale=.21]{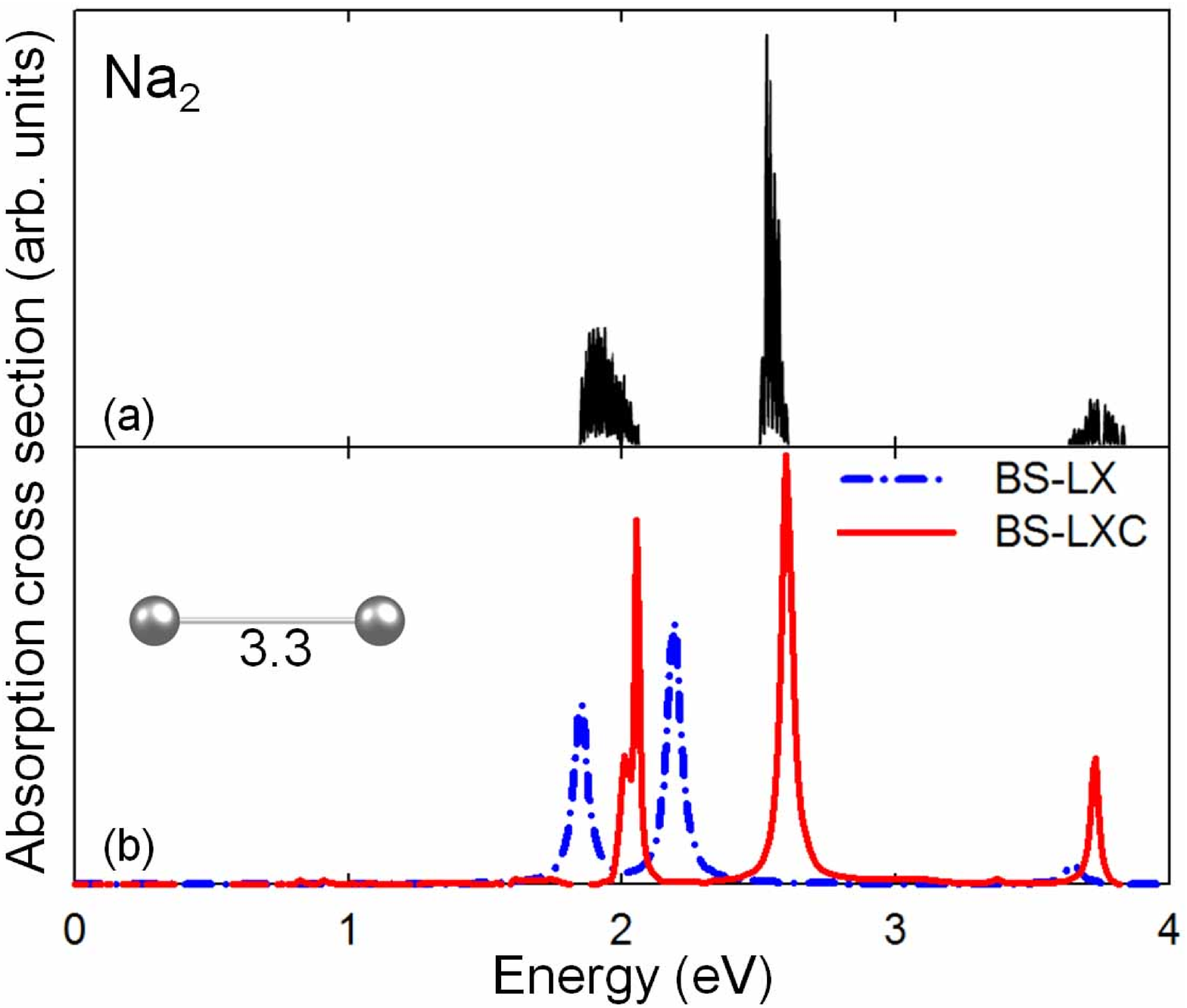}
\caption{Experimental and calculated absorption
spectra of Na$_2$. (a) Experimental spectrum  adapted from
Ref.~\citenum{na2exp}. (b) Cross section calculated from
$\chi^{\mathrm{r}}$ in the BS-LXC (solid line) and BS-LX
(dot-dashed) approximation. The inset shows the cluster structure with
the distance in {\AA}.
 \label{FIGna2}}
\end{figure}

Within the present approach, the computed spectra should fulfil the  $f$-sum
rule for finite systems
\begin{eqnarray}\label{fsumrule}
\int  {\rm d} \omega \, \omega & \sum_{n_1\dots n_4} \frac{1}{3}
\sum_{\mu=x,y,z} d_{\mu,n_1n_2} d_{\mu,n_3n_4}
 \\
& \times {\rm Im} \langle n_2n_1|\chi^{\rm r}(\omega)|n_3n_4
\rangle= -\frac{\hbar^2 \pi}{m}N_e  \nonumber
\end{eqnarray}
for the
conserving BS-LX and BS-LXC approximations, 
Here, $m$ is the electron mass and $N_e$ is the total number of
electrons in the system. Equation \eqref{fsumrule} is equivalent to
the Thomas-Reiche-Kuhn sum rule for the absorption cross section and
is directly related to the particle number conservation
law~\cite{PalEPJB}. For our results, Eq.~\eqref{fsumrule} is
numerically checked and the $f$-sum rule is fulfilled to better than
99\%.

Calculated photoabsorption spectra for Na$_2$ are shown in
Fig.~\ref{FIGna2} and compared to experimental data. Only when
correlations are included in the computation of $\chi^{\mathrm{r}}$,
the theoretical spectrum  reproduces the experimental
absorption peaks which occur around 1.97, 2.55, and 3.75~eV. The
pronounced difference between the  BS-LX and the
BS-LXC results (for the same ground-state cluster geometry which is optimized)
shows the importance of electronic correlation
contributions to the absorption spectra.

\begin{figure}[htb!]
\includegraphics[scale=.21]{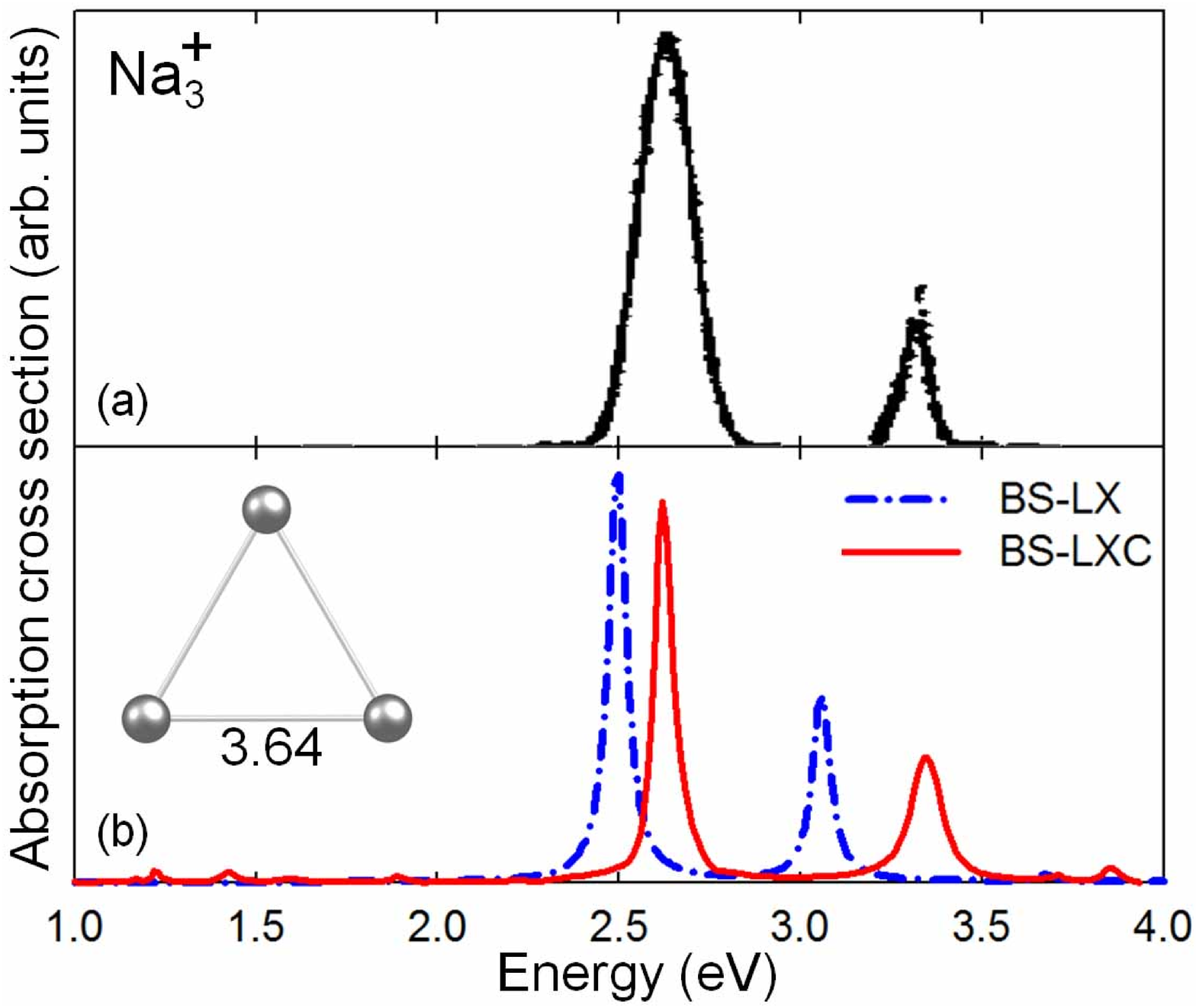}
\caption{Experimental and calculated absorption
spectra of Na$_3^+$. (a) Experimental spectrum  adapted from
Ref.~\citenum{Haberland2Exp}. (b) Cross section is calculated
in the BS-LXC (solid line) and BS-LX
(dot-dashed) approximation. The inset shows the cluster structure with
the distances in {\AA}.
 \label{FIGna3}}
\end{figure}

In Fig.~\ref{FIGna3} calculated and experimental results are shown for  Na$_3^+$. Again,
the theoretical spectrum resolves  both the  larger experimental peak
at 2.62~eV as well as the smaller peak at 3.33~eV only when
correlations are taken into account. Without  correlation contributions, the BS-LX approximation incorrectly describes the
positions of the peaks.

\begin{figure}[htb!]
\includegraphics[scale=.25]{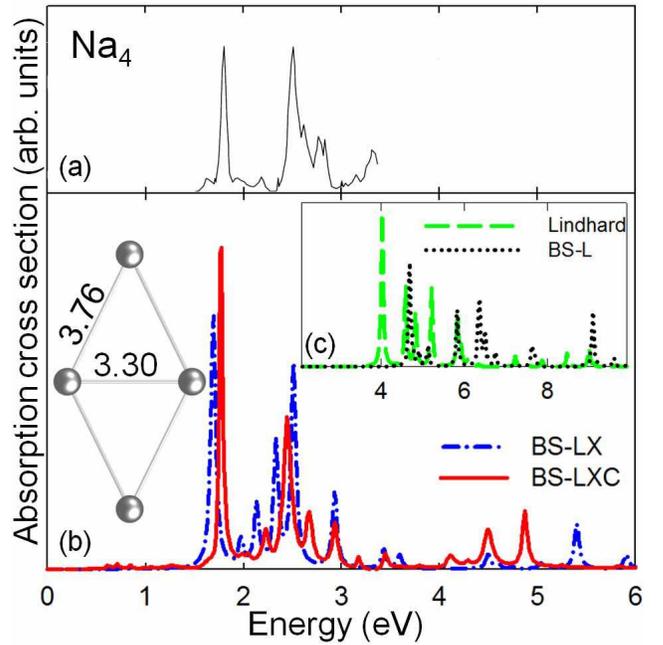}
\caption{Experimental and calculated absorption
spectra of Na$_4$. (a) Experimental spectrum  adapted from
Refs.~\citenum{Kappes1Exp} and~\citenum{Kappes2Exp}. (b) The
cross section obtained from the BS-LXC (solid line) and BS-LX
(dot-dashed) approximation. The inset shows the cluster structure with
the distances in {\AA}. (c) The spectra obtained from the
Lindhard (dashed) and BS-L (dotted) approximations.
 \label{FIGna4}}
\end{figure}

In the case of Na$_4$, the BS-LX and BS-LXC approximations yield
peaks in the absorption cross section that are energetically close, as can be seen from the panel
(b) of Fig.~\ref{FIGna4}. However, the inclusion of correlation
effects results in a better agreement with the
experimental spectrum shown in panel (a) of Fig.~\ref{FIGna4},
for which the main peaks are centered around 1.8 and 2.5~eV. Panel
(c) shows  the calculated spectra corresponding to the
Lindhard and the BS-L approximation for $\chi^{\mathrm{r}}$.  These approximations yield completely different peak positions
that are red shifted with about 2~eV. Moreover, the $f$-sum rule is
not fulfilled for the Lindhard and the BS-L approximations, due to
the inconsistency between the single- and the two-particle
quantities used.

\begin{figure}[htb!]
\includegraphics[scale=.23]{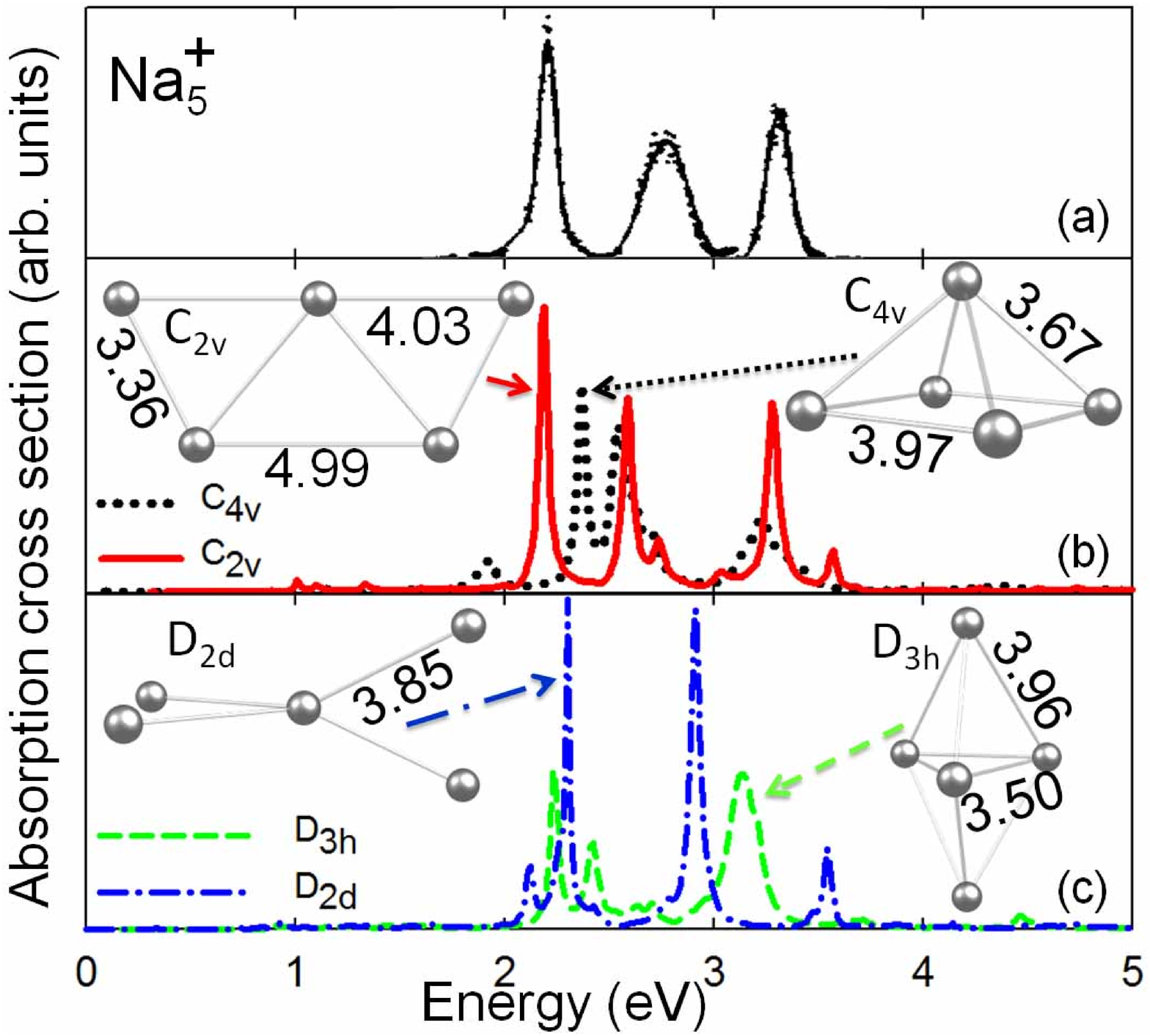}
\caption{Experimental and calculated absorption
spectra of Na$_5^+$. (a) Experimental spectrum  adapted from
Refs.~\citenum{Haberland2Exp} and ~\citenum{Haberland4Exp}.
(b), (c) Cross sections in the BS-LXC
approximation for clusters with different symmetries:
C$_{\mathrm{2v}}$ (solid line), C$_{\mathrm{4v}}$ (dotted),
D$_{\mathrm{2d}}$ (dot-dashed), and D$_{\mathrm{3h}}$ (dashed).
 The insets show the cluster structures with the
distances in {\AA}.
 \label{FIGna5}}
\end{figure}

\begin{table}
 \caption{ Ground-state total energies (in eV) of the Na$_5^+$, Na$_6$, Na$_7^+$, and Na$_8$ clusters
for several optimized geometrical structures (shown in the panels (b) and (c) of
Figs.~\ref{FIGna5}, \ref{FIGna6}, \ref{FIGna7}, and \ref{FIGna8}, respectively). The configuration with the lowest energy
corresponds to the most stable geometry.}
\label{TotalEn}
\begin{tabular}{cccc}
\hline
  Na$_5^+$                   & Na$_6$                      & Na$_7^+$                                   & Na$_8$   \\ \hline
 -21.605 (D$_{\mathrm{2d}}$) & -29.970 (planar)            & -31.866 (D$_{\mathrm{5h}}$)                & -40.121 (D$_{\mathrm{2d}}^{\mathrm{(c)}}$)  \\
 -21.476 (C$_{\mathrm{2v}}$) & -29.821 (C$_{\mathrm{5v}}$) & -31.329 (C$_{\mathrm{2v}}^{\mathrm{(b)}}$) & -39.934 (D$_{\mathrm{4d}}$)  \\
 -21.168 (D$_{\mathrm{3h}}$) & -29.526 (D$_{\mathrm{4h}}$) & -30.984 (C$_{\mathrm{3v}}$)                & -39.661 (D$_{\mathrm{2d}}^{\mathrm{(b)}}$)  \\
 -20.526 (C$_{\mathrm{4v}}$) & -28.937 (C$_{\mathrm{2v}}$) & -30.829 (C$_{\mathrm{2v}}^{\mathrm{(c)}}$) & -39.934 (D$_{\mathrm{4d}}$) \\
\hline
 \end{tabular}
\end{table}

To investigate possible geometries of the clusters that may play a role under typical
 experimental conditions, we consider four different symmetries
for Na$_5^+$: C$_{\mathrm{2v}}$, C$_{\mathrm{4v}}$,
D$_{\mathrm{2d}}$, and D$_{\mathrm{3h}}$. In Fig.~\ref{FIGna5} we
present  spectra calculated in the  BS-LXC approximation for each
symmetry, and compare them to experiment. As can be seen from
the panels (b) and (c) of  Fig.~\ref{FIGna5}, the cluster structures
which yield the best agreement to the measured cross section
correspond to the  D$_{\mathrm{2d}}$ and C$_{\mathrm{2v}}$ symmetries. They best
 resolve  the experimental peaks at 2.2, 2.7, and 3.3~eV. The
D$_{\mathrm{2d}}$ and C$_{\mathrm{2v}}$ have similar total ground state
energies (0.13~eV difference), and are energetically
more stable than the   D$_{\mathrm{3h}}$ and C$_{\mathrm{4v}}$
structures. In an experiment,  at finite temperature,
it is possible that both   the C$_{\mathrm{2v}}$ and the
D$_{\mathrm{2d}}$ clusters are responsible for the shape of the
individual peaks, but a clear distinction between the weight of the
contribution of  each cluster is not possible with a $T=0$\,K
theory.

\begin{figure}[htb!]
\includegraphics[scale=.22]{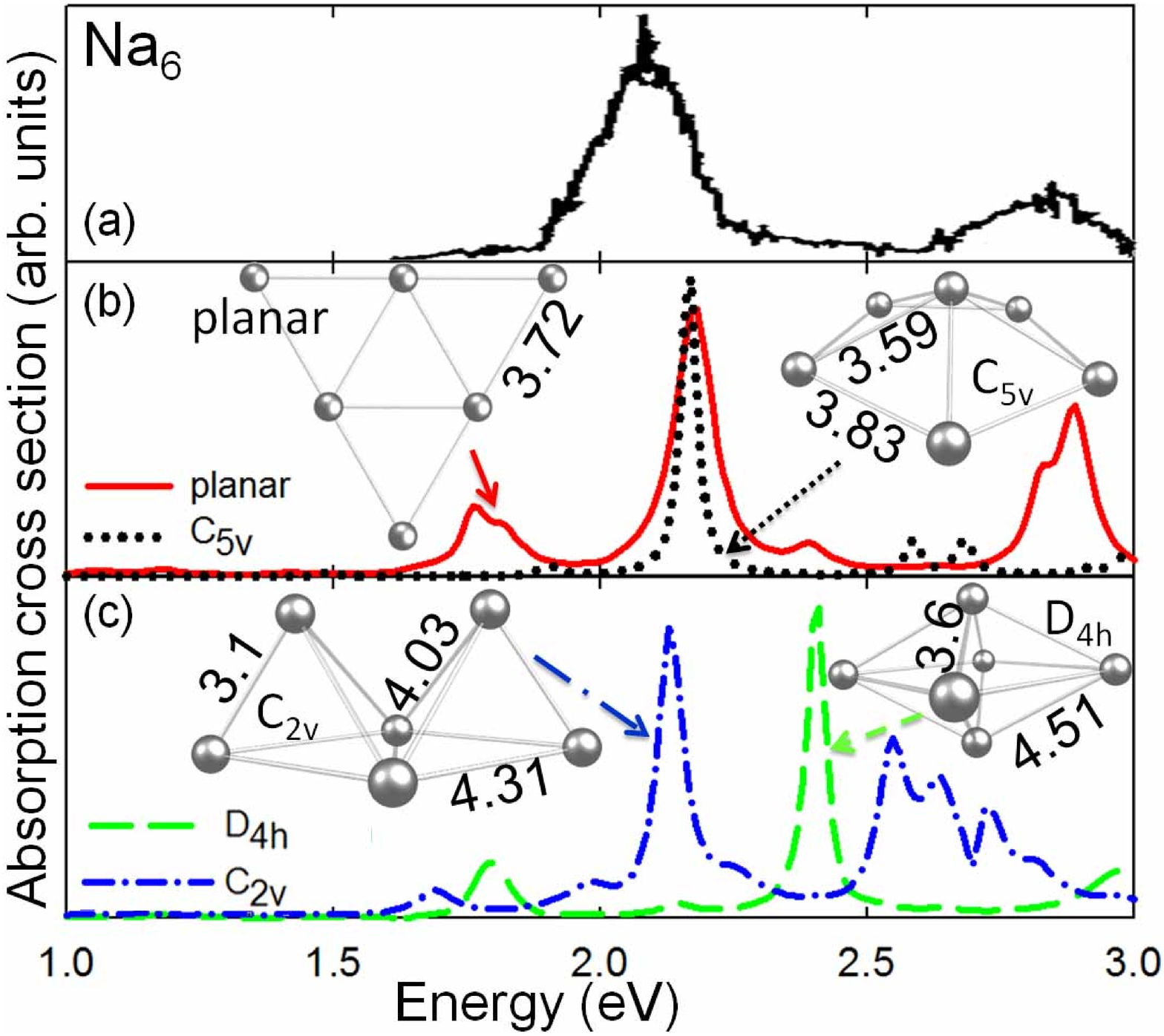}
\caption{Experimental and calculated absorption
spectra of Na$_6$. (a) Experimental spectrum  adapted from
Ref.~\citenum{na6exp}. (b), (c) Cross sections in the BS-LXC approximation for clusters with different symmetries:
planar (solid line), C$_{\mathrm{5v}}$ (dotted), C$_{\mathrm{2v}}$
(dot-dashed), and D$_{\mathrm{4h}}$ (dashed).
 The insets show the cluster structures with the
distances in {\AA}.
 \label{FIGna6}}
\end{figure}

For Na$_6$ we have also investigated four different symmetries of the
cluster structure: planar, C$_{\mathrm{5v}}$, C$_{\mathrm{2v}}$, and
D$_{\mathrm{4h}}$. From Fig.~\ref{FIGna6} one can see that the cross
section of the  planar and the C$_{\mathrm{2v}}$ geometries agree
best with the experimental spectrum. The  energetically most stable
structure is the planar Na$_6$. The planar structure overestimates
by 0.1~eV the position of the dominant experimental peak at
2.1~eV.  It resolves the peak around 2.7~eV, but predicts
another peak at 1.75~eV, which is not present in the experimental
spectrum. In the case of the C$_{\mathrm{2v}}$ cluster, the most prominent
peak agrees well with the experimental peak at 2.1~eV, while the
position of the smaller experimental peak around 2.7~eV is
underestimated by the calculation by 0.2~eV.

\begin{figure}[htb!]
\includegraphics[scale=.23]{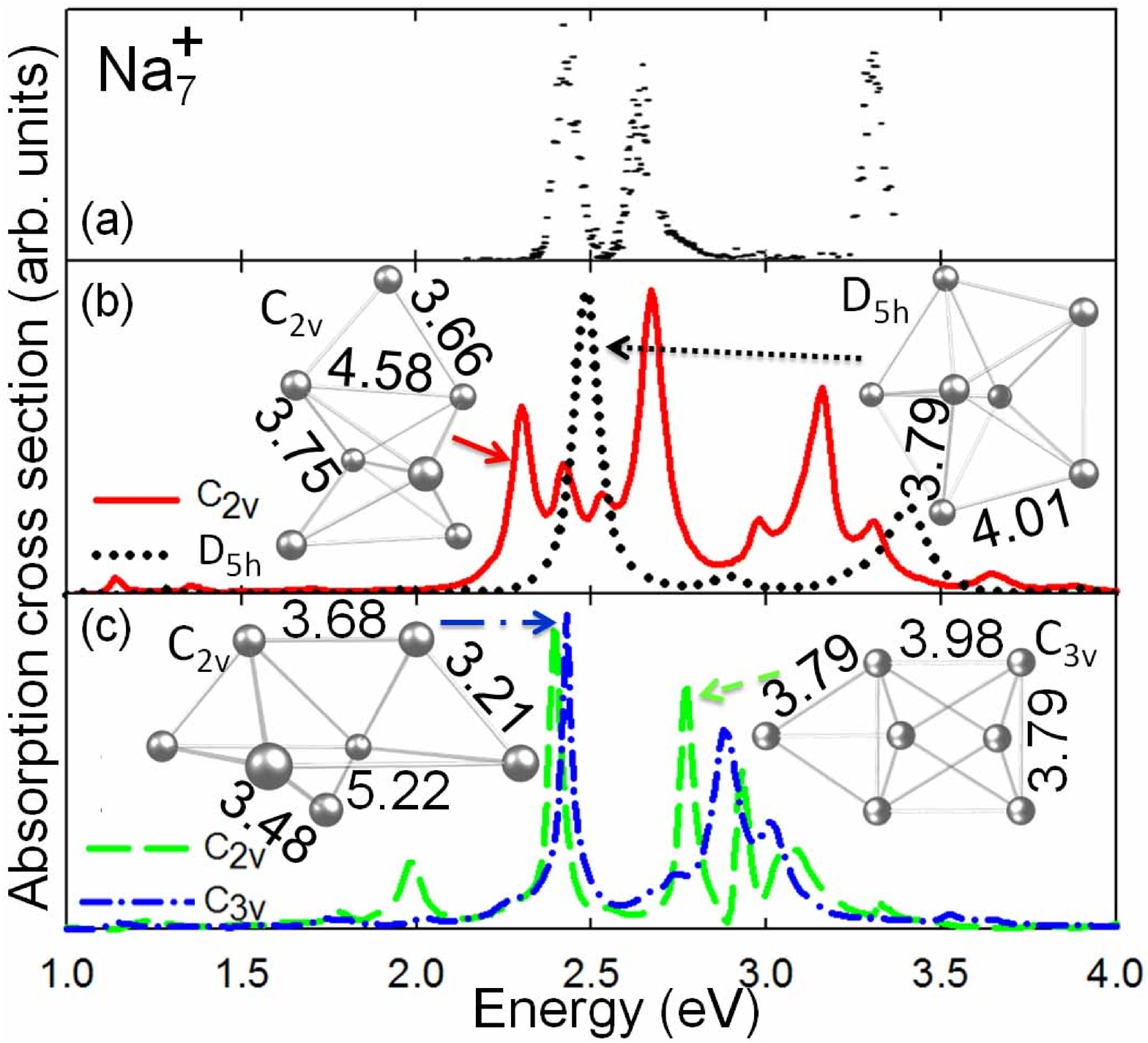}
\caption{Experimental and calculated absorption
spectra of Na$_7^+$. (a) Experimental spectrum  adapted from
Ref.~\citenum{Haberland4Exp}. (b), (c) Cross sections
in the BS-LXC approximation for clusters with
different symmetries: two C$_{\mathrm{2v}}$ (solid and dot-dashed
lines), D$_{\mathrm{5h}}$ (dotted), and C$_{\mathrm{3v}}$ (dashed).
 The insets show the cluster geometries with the
distances in {\AA}.
 \label{FIGna7}}
\end{figure}

In Fig.~\ref{FIGna7} we show a comparison between the experimental
and the theoretical absorption spectra of Na$_7^+$, calculated for
two different clusters of C$_{\mathrm{2v}}$ symmetries and also for
the  D$_{\mathrm{5h}}$ and C$_{\mathrm{3v}}$ geometries. The
structure with the most stable energetic configuration is the
D$_{\mathrm{5h}}$, which also reproduces well the 2.4 and 3.3~eV
experimental peaks, although with a small offset of  0.1~eV. The
experimental peak at 2.4~eV is well reproduced by the structures
from panel (c) of Fig.~\ref{FIGna7}, while the C$_{\mathrm{2v}}$
structure from panel (b) resolves well the experimental peak at 2.65~eV.

\begin{figure}[htb!]
\includegraphics[scale=.23]{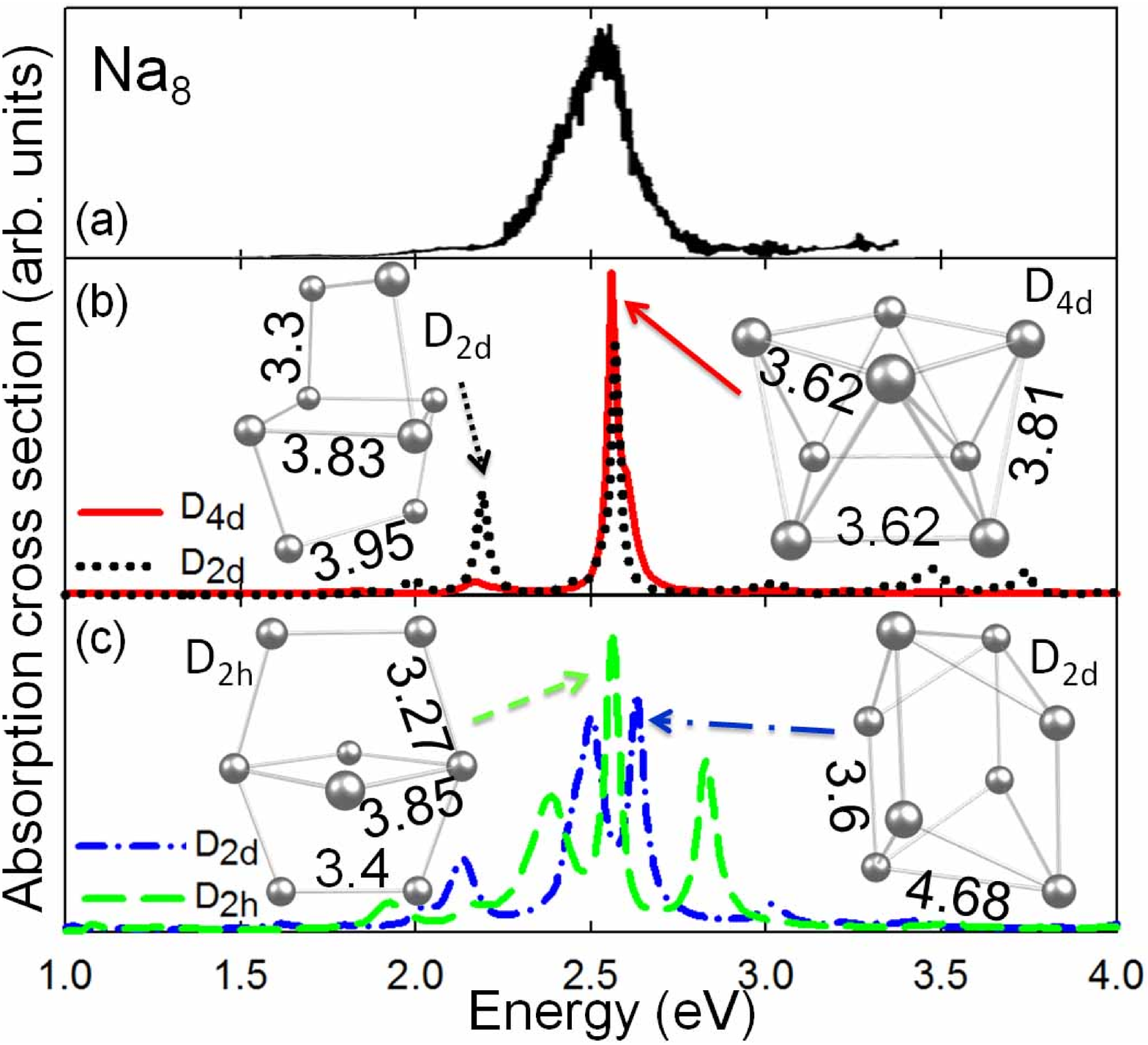}
\caption{(color online) Experimental and calculated absorption
spectra of Na$_8$. (a) Experimental spectrum  adapted from
Refs.~\citenum{Kappes1Exp} and~\citenum{Kappes2Exp}. (b), (c)
Cross sections in the BS-LXC approximation for
clusters with different symmetries: D$_{\mathrm{4d}}$ (solid line),
two D$_{\mathrm{2d}}$ (dotted  and dot-dashed) and D$_{\mathrm{2h}}$
(dashed).
 The insets show the cluster structures with the
distances in {\AA}.
 \label{FIGna8}}
\end{figure}

The measured spectrum of the Na$_8$ cluster in Fig.~\ref{FIGna8} exhibits a single
broad peak extending from  2.2 to 2.8~eV. The
theoretical result is computed for four different structural
symmetries: D$_{\mathrm{4d}}$, D$_{\mathrm{2h}}$, and  two different  D$_{\mathrm{2d}}$ structures,  as shown in Fig.~\ref{FIGna8}. Both the
D$_{\mathrm{4d}}$ and the D$_{\mathrm{2d}}$ structures from panel
(b) give a pronounced  peak around 2.55~eV, which agrees well with the
experimental data. However, the D$_{\mathrm{2d}}$ cluster predicts a
smaller peak at 2.19~eV  not present in the experiment. The
D$_{\mathrm{2d}}$ (which has the lowest ground state energy) and the
D$_{\mathrm{2h}}$ structures from panel (c) yield several
photoabsorption resonance peaks that lie in the same energy
interval as the measured spectra. Therefore, different possible configurations
of the cluster geometries and finite temperature might be
responsible for the strongly broadened experimental line. However, as
mentioned above, it is not possible to assign a  clear
 weight for each individual structure computed for $T=0$\,K. At finite
temperatures  phonons are present and the eigenmodes of these vibronic
states do not only broaden the peaks, but they may also alter the
selection rules, leading to new peaks in the absorption
spectra~\cite{GLyork, GLPRBphonons}.

Summarizing, the calculated BS-LXC spectra presented here are in very good
agreement with the measured photoabsorption  cross sections, given
the potential problems for experiment-theory comparisons which are
due to the unknown cluster structures, finite temperature and
inhomogeneous broadening effects that tend to wash out the intrinsic
spectral features of the resonances.

\section{Conclusions}\label{SectionConclusions}

The absorption spectra of small closed-shell Na clusters were
calculated by means of a linear response approach for the
electron-hole correlation function,  defined as the
functional derivative with respect to a weak external perturbation
of the non-equilibrium single-particle density matrix. We numerically solved  a
 Bethe-Salpeter-like  equation for electron-hole  correlation
that obeys physical conservation laws by construction. In particular, we numerically
checked that the $f$-sum rule is fulfilled to better that 99\%. We found that
correlations effects beyond the mean-field limit need to be  included in the
calculations in order to properly account for the positions and the
widths of the resonance peaks. In the present approach, the finite
broadening of the photoabsorption lines is attributed to the
correlations between the interacting
quasielectrons and holes  with a finite lifetime.

The computed  cross sections of the clusters displayed an overall good
agreement with the measured spectra with respect to the positions
and the shapes of the photoabsorption lines. For the smaller
clusters Na$_2$-Na$_3$ we found the geometrical configurations
that yield theoretical spectra in excellent agreement with the
experiment. For the larger clusters Na$_5^+$-Na$_8$ we have
considered different energetically stable geometries, in order to
account, within a $T=0$\,K theory, for finite-temperature effects
which are unavoidable  in experiments. We were able to pinpoint the cluster
geometries likely present in the experiments, where the structure cannot be directly obtained.

\begin{acknowledgments}
The authors  acknowledge support from the German Research Foundation  through the Priority Programme
1153. The authors thank
Y.~Pavlyukh for stimulating and useful discussions.
\end{acknowledgments}

\nocite{*}


\begin{thebibliography}{10}%
\makeatletter
\providecommand \@ifxundefined [1]{%
 \ifx #1\undefined \expandafter \@firstoftwo
 \else \expandafter \@secondoftwo
\fi
}%
\providecommand \@ifnum [1]{%
 \ifnum #1\expandafter \@firstoftwo
 \else \expandafter \@secondoftwo
\fi
}%
\providecommand \enquote [1]{``#1''}%
\providecommand \bibnamefont  [1]{#1}%
\providecommand \bibfnamefont [1]{#1}%
\providecommand \citenamefont [1]{#1}%
\providecommand\href[0]{\@sanitize\@href}%
\providecommand\@href[1]{\endgroup\@@startlink{#1}\endgroup\@@href}%
\providecommand\@@href[1]{#1\@@endlink}%
\providecommand \@sanitize [0]{\begingroup\catcode`\&12\catcode`\#12\relax}%
\@ifxundefined \pdfoutput {\@firstoftwo}{%
 \@ifnum{\z@=\pdfoutput}{\@firstoftwo}{\@secondoftwo}%
}{%
 \providecommand\@@startlink[1]{\leavevmode}%
 \providecommand\@@endlink[0]{}%
}{%
 \providecommand\@@startlink[1]{%
  \leavevmode
  \pdfstartlink
   attr{/Border[0 0 1 ]/H/I/C[0 1 1]}%
   user{/Subtype/Link/A<</Type/Action/S/URI/URI(#1)>>}%
  \relax
 }%
 \providecommand\@@endlink[0]{\pdfendlink}%
}%
\providecommand \url  [0]{\begingroup\@sanitize \@url }%
\providecommand \@url [1]{\endgroup\@href {#1}{\urlprefix}}%
\providecommand \urlprefix [0]{URL }%
\providecommand \Eprint[0]{\href }%
\@ifxundefined \urlstyle {%
  \providecommand \doi [1]{doi:\discretionary{}{}{}#1}%
}{%
  \providecommand \doi [0]{doi:\discretionary{}{}{}\begingroup
  \urlstyle{rm}\Url }%
}%
\providecommand \doibase [0]{http://dx.doi.org/}%
\providecommand \Doi[1]{\href{\doibase#1}}%
\providecommand \bibAnnote [3]{%
  \BibitemShut{#1}%
  \begin{quotation}\noindent
    \textsc{Key:}\ #2\\\textsc{Annotation:}\ #3%
  \end{quotation}%
}%
\providecommand \bibAnnoteFile [2]{%
  \IfFileExists{#2}{\bibAnnote {#1} {#2} {\input{#2}}}{}%
}%
\providecommand \typeout [0]{\immediate \write \m@ne }%
\providecommand \selectlanguage [0]{\@gobble}%
\providecommand \bibinfo [0]{\@secondoftwo}%
\providecommand \bibfield [0]{\@secondoftwo}%
\providecommand \translation [1]{[#1]}%
\providecommand \BibitemOpen[0]{}%
\providecommand \bibitemStop [0]{}%
\providecommand \bibitemNoStop [0]{.\EOS\space}%
\providecommand \EOS [0]{\spacefactor3000\relax}%
\providecommand \BibitemShut [1]{\csname bibitem#1\endcsname}%
\bibitem{Knight1}%
  \BibitemOpen
  \bibfield{author}{%
  \bibinfo {author} {\bibfnamefont{W.~D.}\ \bibnamefont{Knight}}, \bibinfo
  {author} {\bibfnamefont{K.}~\bibnamefont{Clemenger}}, \bibinfo {author}
  {\bibfnamefont{W.~A.}\ \bibnamefont{de~Heer}}, \bibinfo {author}
  {\bibfnamefont{W.~A.}\ \bibnamefont{Saunders}}, \bibinfo {author}
  {\bibfnamefont{M.~Y.}\ \bibnamefont{Chou}},\ and\ \bibinfo {author}
  {\bibfnamefont{M.~L.}\ \bibnamefont{Cohen}},\ }%
  \bibfield{journal}{%
  \bibinfo {journal} {Phys. Rev. Lett.}\ }%
  \textbf{\bibinfo {volume} {52}},\ \bibinfo {pages} {2141} (\bibinfo {year}
  {1984})%
  \bibAnnoteFile{NoStop}{Knight1}%
\bibitem{Knight2Exp}%
  \BibitemOpen
  \bibfield{author}{%
  \bibinfo {author} {\bibfnamefont{W.~A.}\ \bibnamefont{de~Heer}}, \bibinfo
  {author} {\bibfnamefont{K.}~\bibnamefont{Selby}}, \bibinfo {author}
  {\bibfnamefont{V.}~\bibnamefont{Kresin}}, \bibinfo {author}
  {\bibfnamefont{J.}~\bibnamefont{Masui}}, \bibinfo {author}
  {\bibfnamefont{M.}~\bibnamefont{Vollmer}}, \bibinfo {author}
  {\bibfnamefont{A.}~\bibnamefont{Chatelain}},\ and\ \bibinfo {author}
  {\bibfnamefont{W.~D.}\ \bibnamefont{Knight}},\ }%
  \bibfield{journal}{%
  \bibinfo {journal} {Phys. Rev. Lett.}\ }%
  \textbf{\bibinfo {volume} {59}},\ \bibinfo {pages} {1805} (\bibinfo {year}
  {1987})%
  \bibAnnoteFile{NoStop}{Knight2Exp}%
\bibitem{Kappes1Exp}%
  \BibitemOpen
  \bibfield{author}{%
  \bibinfo {author} {\bibfnamefont{C.~R.~C.}\ \bibnamefont{Wang}}, \bibinfo
  {author} {\bibfnamefont{S.}~\bibnamefont{Pollack}}, \bibinfo {author}
  {\bibfnamefont{D.}~\bibnamefont{Cameron}},\ and\ \bibinfo {author}
  {\bibfnamefont{M.~M.}\ \bibnamefont{Kappes}},\ }%
  \bibfield{journal}{%
  \bibinfo {journal} {J. Chem. Phys.}\ }%
  \textbf{\bibinfo {volume} {93}},\ \bibinfo {pages} {3787} (\bibinfo {year}
  {1990})%
  \bibAnnoteFile{NoStop}{Kappes1Exp}%
\bibitem{Kappes2Exp}%
  \BibitemOpen
  \bibfield{author}{%
  \bibinfo {author} {\bibfnamefont{C.~R.~C.}\ \bibnamefont{Wang}}, \bibinfo
  {author} {\bibfnamefont{S.}~\bibnamefont{Pollack}}, \bibinfo {author}
  {\bibfnamefont{D.}~\bibnamefont{Cameron}},\ and\ \bibinfo {author}
  {\bibfnamefont{M.~M.}\ \bibnamefont{Kappes}},\ }%
  \bibfield{journal}{%
  \bibinfo {journal} {Chem. Phys. Lett.}\ }%
  \textbf{\bibinfo {volume} {166}},\ \bibinfo {pages} {26} (\bibinfo {year}
  {1990})%
  \bibAnnoteFile{NoStop}{Kappes2Exp}%
\bibitem{Knight3Exp}%
  \BibitemOpen
  \bibfield{author}{%
  \bibinfo {author} {\bibfnamefont{K.}~\bibnamefont{Selby}}, \bibinfo {author}
  {\bibfnamefont{V.}~\bibnamefont{Kresin}}, \bibinfo {author}
  {\bibfnamefont{J.}~\bibnamefont{Masui}}, \bibinfo {author}
  {\bibfnamefont{M.}~\bibnamefont{Vollmer}}, \bibinfo {author}
  {\bibfnamefont{W.~A.}\ \bibnamefont{de~Heer}}, \bibinfo {author}
  {\bibfnamefont{A.}~\bibnamefont{Scheidemann}},\ and\ \bibinfo {author}
  {\bibfnamefont{W.~D.}\ \bibnamefont{Knight}},\ }%
  \bibfield{journal}{%
  \bibinfo {journal} {Phys. Rev. B}\ }%
  \textbf{\bibinfo {volume} {43}},\ \bibinfo {pages} {4565} (\bibinfo {year}
  {1991})%
  \bibAnnoteFile{NoStop}{Knight3Exp}%
\bibitem{deHeer}%
  \BibitemOpen
  \bibfield{author}{%
  \bibinfo {author} {\bibfnamefont{W.~A.}\ \bibnamefont{de~Heer}},\ }%
  \bibfield{journal}{%
  \bibinfo {journal} {Rev. Mod. Phys.}\ }%
  \textbf{\bibinfo {volume} {65}},\ \bibinfo {pages} {611} (\bibinfo {year}
  {1993})%
  \bibAnnoteFile{NoStop}{deHeer}%
\bibitem{Haberland3Exp}%
  \BibitemOpen
  \bibfield{author}{%
  \bibinfo {author} {\bibfnamefont{T.}~\bibnamefont{Reiners}}, \bibinfo
  {author} {\bibfnamefont{W.}~\bibnamefont{Orlik}}, \bibinfo {author}
  {\bibfnamefont{C.}~\bibnamefont{Ellert}}, \bibinfo {author}
  {\bibfnamefont{M.}~\bibnamefont{Schmidt}},\ and\ \bibinfo {author}
  {\bibfnamefont{H.}~\bibnamefont{Haberland}},\ }%
  \bibfield{journal}{%
  \bibinfo {journal} {Chem. Phys. Lett.}\ }%
  \textbf{\bibinfo {volume} {215}},\ \bibinfo {pages} {357} (\bibinfo {year}
  {1993})%
  \bibAnnoteFile{NoStop}{Haberland3Exp}%
\bibitem{Haberland1Exp}%
  \BibitemOpen
  \bibfield{author}{%
  \bibinfo {author} {\bibfnamefont{C.}~\bibnamefont{Ellert}}, \bibinfo {author}
  {\bibfnamefont{M.}~\bibnamefont{Schmidt}}, \bibinfo {author}
  {\bibfnamefont{C.}~\bibnamefont{Schmitt}}, \bibinfo {author}
  {\bibfnamefont{T.}~\bibnamefont{Reiners}},\ and\ \bibinfo {author}
  {\bibfnamefont{H.}~\bibnamefont{Haberland}},\ }%
  \bibfield{journal}{%
  \bibinfo {journal} {Phys. Rev. Lett.}\ }%
  \textbf{\bibinfo {volume} {75}},\ \bibinfo {pages} {1731} (\bibinfo {year}
  {1995})%
  \bibAnnoteFile{NoStop}{Haberland1Exp}%
\bibitem{Haberland2Exp}%
  \BibitemOpen
  \bibfield{author}{%
  \bibinfo {author} {\bibfnamefont{M.}~\bibnamefont{Schmidt}}\ and\ \bibinfo
  {author} {\bibfnamefont{H.}~\bibnamefont{Haberland}},\ }%
  \bibfield{journal}{%
  \bibinfo {journal} {Eur. Phys. J. D}\ }%
  \textbf{\bibinfo {volume} {6}},\ \bibinfo {pages} {109} (\bibinfo {year}
  {1999})%
  \bibAnnoteFile{NoStop}{Haberland2Exp}%
\bibitem{Haberland4Exp}%
  \BibitemOpen
  \bibfield{author}{%
  \bibinfo {author} {\bibfnamefont{M.}~\bibnamefont{Schmidt}}, \bibinfo
  {author} {\bibfnamefont{C.}~\bibnamefont{Ellert}}, \bibinfo {author}
  {\bibfnamefont{W.}~\bibnamefont{Kronm\"uller}},\ and\ \bibinfo {author}
  {\bibfnamefont{H.}~\bibnamefont{Haberland}},\ }%
  \bibfield{journal}{%
  \bibinfo {journal} {Phys. Rev. B}\ }%
  \textbf{\bibinfo {volume} {59}},\ \bibinfo {pages} {10970} (\bibinfo {year}
  {1999})%
  \bibAnnoteFile{NoStop}{Haberland4Exp}%
\bibitem{IssendorffExp}%
  \BibitemOpen
  \bibfield{author}{%
  \bibinfo {author} {\bibfnamefont{G.}~\bibnamefont{Wrigge}}, \bibinfo {author}
  {\bibfnamefont{M.~Astruc}\ \bibnamefont{Hoffmann}},\ and\ \bibinfo {author}
  {\bibfnamefont{B.~v.}\ \bibnamefont{Issendorff}},\ }%
  \bibfield{journal}{%
  \bibinfo {journal} {Phys. Rev. A}\ }%
  \textbf{\bibinfo {volume} {65}},\ \bibinfo {pages} {063201} (\bibinfo {year}
  {2002})%
  \bibAnnoteFile{NoStop}{IssendorffExp}%
\bibitem{BalettoExp}%
  \BibitemOpen
  \bibfield{author}{%
  \bibinfo {author} {\bibfnamefont{F.}~\bibnamefont{Baletto}}\ and\ \bibinfo
  {author} {\bibfnamefont{R.}~\bibnamefont{Ferrando}},\ }%
  \bibfield{journal}{%
  \bibinfo {journal} {Rev. Mod. Phys.}\ }%
  \textbf{\bibinfo {volume} {77}},\ \bibinfo {pages} {371} (\bibinfo {year}
  {2005})%
  \bibAnnoteFile{NoStop}{BalettoExp}%
\bibitem{ClemengerShell1985}%
  \BibitemOpen
  \bibfield{author}{%
  \bibinfo {author} {\bibfnamefont{K.}~\bibnamefont{Clemenger}},\ }%
  \bibfield{journal}{%
  \bibinfo {journal} {Phys. Rev. B}\ }%
  \textbf{\bibinfo {volume} {32}},\ \bibinfo {pages} {1359} (\bibinfo {year}
  {1985})%
  \bibAnnoteFile{NoStop}{ClemengerShell1985}%
\bibitem{BeckJellium1984}%
  \BibitemOpen
  \bibfield{author}{%
  \bibinfo {author} {\bibfnamefont{D.~E.}\ \bibnamefont{Beck}},\ }%
  \bibfield{journal}{%
  \bibinfo {journal} {Phys. Rev. B}\ }%
  \textbf{\bibinfo {volume} {30}},\ \bibinfo {pages} {6935} (\bibinfo {year}
  {1984})%
  \bibAnnoteFile{NoStop}{BeckJellium1984}%
\bibitem{EckardtJellium1984}%
  \BibitemOpen
  \bibfield{author}{%
  \bibinfo {author} {\bibfnamefont{W.}~\bibnamefont{Ekardt}},\ }%
  \bibfield{journal}{%
  \bibinfo {journal} {Phys. Rev. Lett.}\ }%
  \textbf{\bibinfo {volume} {52}},\ \bibinfo {pages} {1925} (\bibinfo {year}
  {1984})%
  \bibAnnoteFile{NoStop}{EckardtJellium1984}%
\bibitem{BrackJellium1993}%
  \BibitemOpen
  \bibfield{author}{%
  \bibinfo {author} {\bibfnamefont{M.}~\bibnamefont{Brack}},\ }%
  \bibfield{journal}{%
  \bibinfo {journal} {Rev. Mod. Phys.}\ }%
  \textbf{\bibinfo {volume} {65}},\ \bibinfo {pages} {677} (\bibinfo {year}
  {1993})%
  \bibAnnoteFile{NoStop}{BrackJellium1993}%
\bibitem{ReinhardJellium2000}%
  \BibitemOpen
  \bibfield{author}{%
  \bibinfo {author} {\bibfnamefont{S.}~\bibnamefont{K\"ummel}}, \bibinfo
  {author} {\bibfnamefont{M.}~\bibnamefont{Brack}},\ and\ \bibinfo {author}
  {\bibfnamefont{P.-G.}\ \bibnamefont{Reinhard}},\ }%
  \bibfield{journal}{%
  \bibinfo {journal} {Phys. Rev. B}\ }%
  \textbf{\bibinfo {volume} {62}},\ \bibinfo {pages} {7602} (\bibinfo {year}
  {2000})%
  \bibAnnoteFile{NoStop}{ReinhardJellium2000}%
\bibitem{BortignonExp}%
  \BibitemOpen
  \bibfield{author}{%
  \bibinfo {author} {\bibfnamefont{C.}~\bibnamefont{Yannouleas}}, \bibinfo
  {author} {\bibfnamefont{R.~A.}\ \bibnamefont{Broglia}}, \bibinfo {author}
  {\bibfnamefont{M.}~\bibnamefont{Brack}},\ and\ \bibinfo {author}
  {\bibfnamefont{P.~F.}\ \bibnamefont{Bortignon}},\ }%
  \bibfield{journal}{%
  \bibinfo {journal} {Phys. Rev. Lett.}\ }%
  \textbf{\bibinfo {volume} {63}},\ \bibinfo {pages} {255}%
  \bibAnnoteFile{NoStop}{BortignonExp}%
\bibitem{Koutecky31988}%
  \BibitemOpen
  \bibfield{author}{%
  \bibinfo {author}
  {\bibfnamefont{V.}~\bibnamefont{Bona\v{c}i\'{c}-Kouteck\'{y}}}, \bibinfo
  {author} {\bibfnamefont{P.}~\bibnamefont{Fantucci}},\ and\ \bibinfo {author}
  {\bibfnamefont{J.}~\bibnamefont{Kouteck\'{y}}},\ }%
  \bibfield{journal}{%
  \bibinfo {journal} {Phys. Rev. B}\ }%
  \textbf{\bibinfo {volume} {37}},\ \bibinfo {pages} {4369} (\bibinfo {year}
  {1988})%
  \bibAnnoteFile{NoStop}{Koutecky31988}%
\bibitem{Koutecky11990}%
  \BibitemOpen
  \bibfield{author}{%
  \bibinfo {author}
  {\bibfnamefont{V.}~\bibnamefont{Bona\v{c}i\'{c}-Kouteck\'{y}}}, \bibinfo
  {author} {\bibfnamefont{P.}~\bibnamefont{Fantucci}},\ and\ \bibinfo {author}
  {\bibfnamefont{J.}~\bibnamefont{Kouteck\'{y}}},\ }%
  \bibfield{journal}{%
  \bibinfo {journal} {Chem. Phys. Lett.}\ }%
  \textbf{\bibinfo {volume} {166}},\ \bibinfo {pages} {32} (\bibinfo {year}
  {1990})%
  \bibAnnoteFile{NoStop}{Koutecky11990}%
\bibitem{Koutecky21990}%
  \BibitemOpen
  \bibfield{author}{%
  \bibinfo {author}
  {\bibfnamefont{V.}~\bibnamefont{Bona\v{c}i\'{c}-Kouteck\'{y}}}, \bibinfo
  {author} {\bibfnamefont{P.}~\bibnamefont{Fantucci}},\ and\ \bibinfo {author}
  {\bibfnamefont{J.}~\bibnamefont{Kouteck\'{y}}},\ }%
  \bibfield{journal}{%
  \bibinfo {journal} {J. Chem. Phys.}\ }%
  \textbf{\bibinfo {volume} {93}},\ \bibinfo {pages} {3802} (\bibinfo {year}
  {1990})%
  \bibAnnoteFile{NoStop}{Koutecky21990}%
\bibitem{Koutecky41992}%
  \BibitemOpen
  \bibfield{author}{%
  \bibinfo {author}
  {\bibfnamefont{V.}~\bibnamefont{Bona\v{c}i\'{c}-Kouteck\'{y}}}, \bibinfo
  {author} {\bibfnamefont{J.}~\bibnamefont{Pittner}}, \bibinfo {author}
  {\bibfnamefont{C.}~\bibnamefont{Scheuch}}, \bibinfo {author}
  {\bibfnamefont{M.~F.}\ \bibnamefont{Guest}},\ and\ \bibinfo {author}
  {\bibfnamefont{J.}~\bibnamefont{Kouteck\'{y}}},\ }%
  \bibfield{journal}{%
  \bibinfo {journal} {J. Chem. Phys.}\ }%
  \textbf{\bibinfo {volume} {96}},\ \bibinfo {pages} {7938} (\bibinfo {year}
  {1992})%
  \bibAnnoteFile{NoStop}{Koutecky41992}%
\bibitem{Koutecky51996}%
  \BibitemOpen
  \bibfield{author}{%
  \bibinfo {author}
  {\bibfnamefont{V.}~\bibnamefont{Bona\v{c}i\'{c}-Kouteck\'{y}}}, \bibinfo
  {author} {\bibfnamefont{J.}~\bibnamefont{Pittner}}, \bibinfo {author}
  {\bibfnamefont{C.}~\bibnamefont{Fuchs}}, \bibinfo {author}
  {\bibfnamefont{P.}~\bibnamefont{Fantucci}}, \bibinfo {author}
  {\bibfnamefont{M.~F.}\ \bibnamefont{Guest}},\ and\ \bibinfo {author}
  {\bibfnamefont{J.}~\bibnamefont{Kouteck\'{y}}},\ }%
  \bibfield{journal}{%
  \bibinfo {journal} {J. Chem. Phys.}\ }%
  \textbf{\bibinfo {volume} {104}},\ \bibinfo {pages} {1427} (\bibinfo {year}
  {1996})%
  \bibAnnoteFile{NoStop}{Koutecky51996}%
\bibitem{RubioPRL1996}%
  \BibitemOpen
  \bibfield{author}{%
  \bibinfo {author} {\bibfnamefont{A.}~\bibnamefont{Rubio}}, \bibinfo {author}
  {\bibfnamefont{J.~A.}\ \bibnamefont{Alonso}}, \bibinfo {author}
  {\bibfnamefont{X.}~\bibnamefont{Blase}}, \bibinfo {author}
  {\bibfnamefont{L.~C.}\ \bibnamefont{Balb\'as}},\ and\ \bibinfo {author}
  {\bibfnamefont{S.~G.}\ \bibnamefont{Louie}},\ }%
  \bibfield{journal}{%
  \bibinfo {journal} {Phys. Rev. Lett.}\ }%
  \textbf{\bibinfo {volume} {77}},\ \bibinfo {pages} {247} (\bibinfo {year}
  {1996})%
  \bibAnnoteFile{NoStop}{RubioPRL1996}%
\bibitem{OgutPRL1999}%
  \BibitemOpen
  \bibfield{author}{%
  \bibinfo {author} {\bibfnamefont{I.}~\bibnamefont{Vasiliev}}, \bibinfo
  {author} {\bibfnamefont{S.}~\bibnamefont{\"O\ifmmode~\breve{g}\else
  \u{g}\fi{}\"ut}},\ and\ \bibinfo {author} {\bibfnamefont{J.~R.}\
  \bibnamefont{Chelikowsky}},\ }%
  \bibfield{journal}{%
  \bibinfo {journal} {Phys. Rev. Lett.}\ }%
  \textbf{\bibinfo {volume} {82}},\ \bibinfo {pages} {1919} (\bibinfo {year}
  {1999})%
  \bibAnnoteFile{NoStop}{OgutPRL1999}%
\bibitem{RubioJCP2001}%
  \BibitemOpen
  \bibfield{author}{%
  \bibinfo {author} {\bibfnamefont{M.~A.~L.}\ \bibnamefont{Marques}}, \bibinfo
  {author} {\bibfnamefont{A.}~\bibnamefont{Castro}},\ and\ \bibinfo {author}
  {\bibfnamefont{A.}~\bibnamefont{Rubio}},\ }%
  \bibfield{journal}{%
  \bibinfo {journal} {J. Chem. Phys.}\ }%
  \textbf{\bibinfo {volume} {115}},\ \bibinfo {pages} {3006} (\bibinfo {year}
  {2001})%
  \bibAnnoteFile{NoStop}{RubioJCP2001}%
\bibitem{UziPRL2001}%
  \BibitemOpen
  \bibfield{author}{%
  \bibinfo {author} {\bibfnamefont{M.}~\bibnamefont{Moseler}}, \bibinfo
  {author} {\bibfnamefont{H.}~\bibnamefont{H\"akkinen}},\ and\ \bibinfo
  {author} {\bibfnamefont{U.}~\bibnamefont{Landman}},\ }%
  \bibfield{journal}{%
  \bibinfo {journal} {Phys. Rev. Lett.}\ }%
  \textbf{\bibinfo {volume} {87}},\ \bibinfo {pages} {053401} (\bibinfo {year}
  {2001})%
  \bibAnnoteFile{NoStop}{UziPRL2001}%
\bibitem{Broglia2005}%
  \BibitemOpen
  \bibfield{author}{%
  \bibinfo {author} {\bibfnamefont{A.}~\bibnamefont{Fortini}}, \bibinfo
  {author} {\bibfnamefont{M.}~\bibnamefont{Mazzola}}, \bibinfo {author}
  {\bibfnamefont{A.}~\bibnamefont{Mina}}, \bibinfo {author}
  {\bibfnamefont{D.}~\bibnamefont{Provasi}}, \bibinfo {author}
  {\bibfnamefont{G.}~\bibnamefont{Colo}}, \bibinfo {author}
  {\bibfnamefont{G.}~\bibnamefont{Onida}}, \bibinfo {author}
  {\bibfnamefont{H.~E.}\ \bibnamefont{Roman}},\ and\ \bibinfo {author}
  {\bibfnamefont{R.~A.}\ \bibnamefont{Broglia}},\ }%
  \bibfield{journal}{%
  \bibinfo {journal} {J. Phys. B}\ }%
  \textbf{\bibinfo {volume} {38}},\ \bibinfo {pages} {1581} (\bibinfo {year}
  {2005})%
  \bibAnnoteFile{NoStop}{Broglia2005}%
\bibitem{NieminenDFT2008}%
  \BibitemOpen
  \bibfield{author}{%
  \bibinfo {author} {\bibfnamefont{J.-O.}\ \bibnamefont{Joswig}}, \bibinfo
  {author} {\bibfnamefont{L.~O.}\ \bibnamefont{Tunturivuori}},\ and\ \bibinfo
  {author} {\bibfnamefont{R.~M.}\ \bibnamefont{Nieminen}},\ }%
  \bibfield{journal}{%
  \bibinfo {journal} {J. Chem. Phys.}\ }%
  \textbf{\bibinfo {volume} {128}},\ \bibinfo {pages} {14707} (\bibinfo {year}
  {2008})%
  \bibAnnoteFile{NoStop}{NieminenDFT2008}%
\bibitem{OnidaNa4PRL1995}%
  \BibitemOpen
  \bibfield{author}{%
  \bibinfo {author} {\bibfnamefont{G.}~\bibnamefont{Onida}}, \bibinfo {author}
  {\bibfnamefont{L.}~\bibnamefont{Reining}}, \bibinfo {author}
  {\bibfnamefont{R.~W.}\ \bibnamefont{Godby}}, \bibinfo {author}
  {\bibfnamefont{R.}~\bibnamefont{Del~Sole}},\ and\ \bibinfo {author}
  {\bibfnamefont{W.}~\bibnamefont{Andreoni}},\ }%
  \bibfield{journal}{%
  \bibinfo {journal} {Phys. Rev. Lett.}\ }%
  \textbf{\bibinfo {volume} {75}},\ \bibinfo {pages} {818} (\bibinfo {year}
  {1995})%
  \bibAnnoteFile{NoStop}{OnidaNa4PRL1995}%
\bibitem{Louie1PRB2000}%
  \BibitemOpen
  \bibfield{author}{%
  \bibinfo {author} {\bibfnamefont{M.}~\bibnamefont{Rohlfing}}\ and\ \bibinfo
  {author} {\bibfnamefont{S.~G.}\ \bibnamefont{Louie}},\ }%
  \bibfield{journal}{%
  \bibinfo {journal} {Phys. Rev. B}\ }%
  \textbf{\bibinfo {volume} {62}},\ \bibinfo {pages} {4927} (\bibinfo {year}
  {2000})%
  \bibAnnoteFile{NoStop}{Louie1PRB2000}%
\bibitem{OnidaRMP2002}%
  \BibitemOpen
  \bibfield{author}{%
  \bibinfo {author} {\bibfnamefont{G.}~\bibnamefont{Onida}}, \bibinfo {author}
  {\bibfnamefont{L.}~\bibnamefont{Reining}},\ and\ \bibinfo {author}
  {\bibfnamefont{A.}~\bibnamefont{Rubio}},\ }%
  \bibfield{journal}{%
  \bibinfo {journal} {Rev. Mod. Phys.}\ }%
  \textbf{\bibinfo {volume} {74}},\ \bibinfo {pages} {601} (\bibinfo {year}
  {2002})%
  \bibAnnoteFile{NoStop}{OnidaRMP2002}%
\bibitem{LopezPRB2008}%
  \BibitemOpen
  \bibfield{author}{%
  \bibinfo {author} {\bibfnamefont{M.~L.}\ \bibnamefont{del Puerto}}, \bibinfo
  {author} {\bibfnamefont{M.~L.}\ \bibnamefont{Tiago}},\ and\ \bibinfo {author}
  {\bibfnamefont{J.~R.}\ \bibnamefont{Chelikowsky}},\ }%
  \bibfield{journal}{%
  \bibinfo {journal} {Phys. Rev. B}\ }%
  \textbf{\bibinfo {volume} {77}} (\bibinfo {year} {2008})%
  \bibAnnoteFile{NoStop}{LopezPRB2008}%
\bibitem{Ogut2PRB2009}%
  \BibitemOpen
  \bibfield{author}{%
  \bibinfo {author} {\bibfnamefont{M.~L.}\ \bibnamefont{Tiago}}, \bibinfo
  {author} {\bibfnamefont{J.~C.}\ \bibnamefont{Idrobo}}, \bibinfo {author}
  {\bibfnamefont{S.}~\bibnamefont{\"{O}\u{g}\"{u}t}}, \bibinfo {author}
  {\bibfnamefont{J.}~\bibnamefont{Jellinek}},\ and\ \bibinfo {author}
  {\bibfnamefont{J.~R.}\ \bibnamefont{Chelikowsky}},\ }%
  \bibfield{journal}{%
  \bibinfo {journal} {Phys. Rev. B}\ }%
  \textbf{\bibinfo {volume} {79}},\ \bibinfo {pages} {155419} (\bibinfo {year}
  {2009})%
  \bibAnnoteFile{NoStop}{Ogut2PRB2009}%
\bibitem{HedinGW}%
  \BibitemOpen
  \bibfield{author}{%
  \bibinfo {author} {\bibfnamefont{L.}~\bibnamefont{Hedin}},\ }%
  \bibfield{journal}{%
  \bibinfo {journal} {Phys. Rev.}\ }%
  \textbf{\bibinfo {volume} {139}},\ \bibinfo {pages} {A796} (\bibinfo {year}
  {1965})%
  \bibAnnoteFile{NoStop}{HedinGW}%
\bibitem{LouieNaGW}%
  \BibitemOpen
  \bibfield{author}{%
  \bibinfo {author} {\bibfnamefont{S.}~\bibnamefont{Ishii}}, \bibinfo {author}
  {\bibfnamefont{K.}~\bibnamefont{Ohno}}, \bibinfo {author}
  {\bibfnamefont{Y.}~\bibnamefont{Kawazoe}},\ and\ \bibinfo {author}
  {\bibfnamefont{S.~G.}\ \bibnamefont{Louie}},\ }%
  \bibfield{journal}{%
  \bibinfo {journal} {Phys. Rev. B}\ }%
  \textbf{\bibinfo {volume} {63}},\ \bibinfo {pages} {155104} (\bibinfo {year}
  {2001})%
  \bibAnnoteFile{NoStop}{LouieNaGW}%
\bibitem{BaymKadanoff1961}%
  \BibitemOpen
  \bibfield{author}{%
  \bibinfo {author} {\bibfnamefont{G.}~\bibnamefont{Baym}}\ and\ \bibinfo
  {author} {\bibfnamefont{L.~P.}\ \bibnamefont{Kadanoff}},\ }%
  \bibfield{journal}{%
  \bibinfo {journal} {Phys. Rev.}\ }%
  \textbf{\bibinfo {volume} {124}},\ \bibinfo {pages} {287} (\bibinfo {year}
  {1961})%
  \bibAnnoteFile{NoStop}{BaymKadanoff1961}%
\bibitem{PalEPJB}%
  \BibitemOpen
  \bibfield{author}{%
  \bibinfo {author} {\bibfnamefont{G.}~\bibnamefont{Pal}}, \bibinfo {author}
  {\bibfnamefont{Y.}~\bibnamefont{Pavlyukh}}, \bibinfo {author}
  {\bibfnamefont{H.~C.}\ \bibnamefont{Schneider}},\ and\ \bibinfo {author}
  {\bibfnamefont{W.}~\bibnamefont{H\"{u}bner}},\ }%
  \bibfield{journal}{%
  \bibinfo {journal} {Eur. Phys. J. B}\ }%
  \textbf{\bibinfo {volume} {70}},\ \bibinfo {pages} {483} (\bibinfo {year}
  {2009})%
  \bibAnnoteFile{NoStop}{PalEPJB}%
\bibitem{PalAbs1}%
  \BibitemOpen
  \bibfield{author}{%
  \bibinfo {author} {\bibfnamefont{G.}~\bibnamefont{Pal}}, \bibinfo {author}
  {\bibfnamefont{Y.}~\bibnamefont{Pavlyukh}}, \bibinfo {author}
  {\bibfnamefont{W.}~\bibnamefont{H\"{u}bner}},\ and\ \bibinfo {author}
  {\bibfnamefont{H.~C.}\ \bibnamefont{Schneider}},\ }%
  \bibfield{journal}{%
  \bibinfo {journal} {unpublished}}%
   (\bibinfo {year} {2009})%
  \bibAnnoteFile{NoStop}{PalAbs1}%
\bibitem{YaroslavPRL}%
  \BibitemOpen
  \bibfield{author}{%
  \bibinfo {author} {\bibfnamefont{Y.}~\bibnamefont{Pavlyukh}}, \bibinfo
  {author} {\bibfnamefont{J.}~\bibnamefont{Berakdar}},\ and\ \bibinfo {author}
  {\bibfnamefont{W.}~\bibnamefont{H\"{u}bner}},\ }%
  \bibfield{journal}{%
  \bibinfo {journal} {Phys. Rev. Lett.}\ }%
  \textbf{\bibinfo {volume} {100}},\ \bibinfo {pages} {116103} (\bibinfo {year}
  {2008})%
  \bibAnnoteFile{NoStop}{YaroslavPRL}%
\bibitem{YaroslavGW}%
  \BibitemOpen
  \bibfield{author}{%
  \bibinfo {author} {\bibfnamefont{Y.}~\bibnamefont{Pavlyukh}}\ and\ \bibinfo
  {author} {\bibfnamefont{W.}~\bibnamefont{H\"{u}bner}},\ }%
  \bibfield{journal}{%
  \bibinfo {journal} {Phys. Lett. A}\ }%
  \textbf{\bibinfo {volume} {327}},\ \bibinfo {pages} {241} (\bibinfo {year}
  {2004})%
  \bibAnnoteFile{NoStop}{YaroslavGW}%
\bibitem{LANL2DZ}%
  \BibitemOpen
  \bibfield{author}{%
  \bibinfo {author} {\bibfnamefont{P.~J.}\ \bibnamefont{Hay}}\ and\ \bibinfo
  {author} {\bibfnamefont{W.~R.}\ \bibnamefont{Wadt}},\ }%
  \bibfield{journal}{%
  \bibinfo {journal} {J. Chem. Phys.}\ }%
  \textbf{\bibinfo {volume} {82}},\ \bibinfo {pages} {270} (\bibinfo {year}
  {1985})%
  \bibAnnoteFile{NoStop}{LANL2DZ}%
\bibitem{SHC}%
  \BibitemOpen
  \bibfield{author}{%
  \bibinfo {author} {\bibfnamefont{A.~K.}\ \bibnamefont{Rapp\'{e}}}, \bibinfo
  {author} {\bibfnamefont{T.}~\bibnamefont{Smedly}},\ and\ \bibinfo {author}
  {\bibfnamefont{W.~A.}\ \bibnamefont{Goddard~III}},\ }%
  \bibfield{journal}{%
  \bibinfo {journal} {J. Phys. Chem.}\ }%
  \textbf{\bibinfo {volume} {85}},\ \bibinfo {pages} {1662} (\bibinfo {year}
  {1981})%
  \bibAnnoteFile{NoStop}{SHC}%
\bibitem{CEP}%
  \BibitemOpen
  \bibfield{author}{%
  \bibinfo {author} {\bibfnamefont{W.}~\bibnamefont{Stevens}}, \bibinfo
  {author} {\bibfnamefont{H.}~\bibnamefont{Basch}},\ and\ \bibinfo {author}
  {\bibfnamefont{J.}~\bibnamefont{Krauss}},\ }%
  \bibfield{journal}{%
  \bibinfo {journal} {J. Chem. Phys.}\ }%
  \textbf{\bibinfo {volume} {81}},\ \bibinfo {pages} {6026} (\bibinfo {year}
  {1984})%
  \bibAnnoteFile{NoStop}{CEP}%
\bibitem{SDDAll}%
  \BibitemOpen
  \bibfield{author}{%
  \bibinfo {author} {\bibfnamefont{P.}~\bibnamefont{Fuentealba}}, \bibinfo
  {author} {\bibfnamefont{H.}~\bibnamefont{Preuss}}, \bibinfo {author}
  {\bibfnamefont{H.}~\bibnamefont{Stoll}},\ and\ \bibinfo {author}
  {\bibfnamefont{L.}~\bibnamefont{v.~Szentpaly}},\ }%
  \bibfield{journal}{%
  \bibinfo {journal} {Chem. Phys. Lett.}\ }%
  \textbf{\bibinfo {volume} {89}},\ \bibinfo {pages} {418} (\bibinfo {year}
  {1989})%
  \bibAnnoteFile{NoStop}{SDDAll}%
\bibitem{na2exp}%
  \BibitemOpen
  \bibfield{author}{%
  \bibinfo {author} {\bibfnamefont{W.~R.}\ \bibnamefont{Fredrickson}}\ and\
  \bibinfo {author} {\bibfnamefont{William~W.}\ \bibnamefont{Watson}},\ }%
  \bibfield{journal}{%
  \bibinfo {journal} {Phys. Rev.}\ }%
  \textbf{\bibinfo {volume} {30}},\ \bibinfo {pages} {429} (\bibinfo {year}
  {1927})%
  \bibAnnoteFile{NoStop}{na2exp}%
\bibitem{na6exp}%
  \BibitemOpen
  \bibfield{author}{%
  \bibinfo {author} {\bibfnamefont{C.~R.~C.}\ \bibnamefont{Wang}}, \bibinfo
  {author} {\bibfnamefont{S.}~\bibnamefont{Pollack}}, \bibinfo {author}
  {\bibfnamefont{T.~A.}\ \bibnamefont{Dahlseid}}, \bibinfo {author}
  {\bibfnamefont{G.~M.}\ \bibnamefont{Koretsky}},\ and\ \bibinfo {author}
  {\bibfnamefont{M.~M.}\ \bibnamefont{Kappes}},\ }%
  \bibfield{journal}{%
  \bibinfo {journal} {J. Chem. Phys.}\ }%
  \textbf{\bibinfo {volume} {96}},\ \bibinfo {pages} {7931} (\bibinfo {year}
  {1992})%
  \bibAnnoteFile{NoStop}{na6exp}%
\bibitem{GLyork}%
  \BibitemOpen
  \bibfield{author}{%
  \bibinfo {author} {\bibfnamefont{G.}~\bibnamefont{Lefkidis}}, \bibinfo
  {author} {\bibfnamefont{O.}~\bibnamefont{Ney}},\ and\ \bibinfo {author}
  {\bibfnamefont{W.}~\bibnamefont{H\"{u}bner}},\ }%
  \bibfield{journal}{%
  \bibinfo {journal} {phys. stat. solidi (c)}\ }%
  \textbf{\bibinfo {volume} {12}},\ \bibinfo {pages} {4022} (\bibinfo {year}
  {2005})%
  \bibAnnoteFile{NoStop}{GLyork}%
\bibitem{GLPRBphonons}%
  \BibitemOpen
  \bibfield{author}{%
  \bibinfo {author} {\bibfnamefont{G.}~\bibnamefont{Lefkidis}}\ and\ \bibinfo
  {author} {\bibfnamefont{W.}~\bibnamefont{H\"{u}bner}},\ }%
  \bibfield{journal}{%
  \bibinfo {journal} {Phys. Rev. B}\ }%
  \textbf{\bibinfo {volume} {74}},\ \bibinfo {pages} {155106} (\bibinfo {year}
  {2006})%
  \bibAnnoteFile{NoStop}{GLPRBphonons}%
\bibitem{LeuweenPRL}%
  \BibitemOpen
  \bibfield{author}{%
  \bibinfo {author} {\bibfnamefont{N.~E.}\ \bibnamefont{Dahlen}}\ and\ \bibinfo
  {author} {\bibfnamefont{R.}~\bibnamefont{van Leeuwen}},\ }%
  \bibfield{journal}{%
  \bibinfo {journal} {Phys. Rev. Lett.}\ }%
  \textbf{\bibinfo {volume} {98}},\ \bibinfo {pages} {153004} (\bibinfo {year}
  {2007})%
  \bibAnnoteFile{NoStop}{LeuweenPRL}%
\bibitem{MolecContats}%
  \BibitemOpen
  \bibfield{author}{%
  \bibinfo {author} {\bibfnamefont{K.~S.}\ \bibnamefont{Thygesen}}\ and\
  \bibinfo {author} {\bibfnamefont{A.}~\bibnamefont{Rubio}},\ }%
  \bibfield{journal}{%
  \bibinfo {journal} {Phys. Rev. B}\ }%
  \textbf{\bibinfo {volume} {77}},\ \bibinfo {pages} {115333} (\bibinfo {year}
  {2008})%
  \bibAnnoteFile{NoStop}{MolecContats}%
\bibitem{FluctDiss}%
  \BibitemOpen
  \bibfield{author}{%
  \bibinfo {author} {\bibfnamefont{J.}~\bibnamefont{Rammer}}\ and\ \bibinfo
  {author} {\bibfnamefont{H.}~\bibnamefont{Smith}},\ }%
  \bibfield{journal}{%
  \bibinfo {journal} {Rev. Mod. Phys.}\ }%
  \textbf{\bibinfo {volume} {58}},\ \bibinfo {pages} {323} (\bibinfo {year}
  {1986})%
  \bibAnnoteFile{NoStop}{FluctDiss}%
\bibitem{Lipavsky}%
  \BibitemOpen
  \bibfield{author}{%
  \bibinfo {author} {\bibfnamefont{P.}~\bibnamefont{Lipavsk\'y}}, \bibinfo
  {author} {\bibfnamefont{V.}~\bibnamefont{\ifmmode \check{S}\else
  \v{S}\fi{}pi\ifmmode~\check{c}\else \v{c}\fi{}ka}},\ and\ \bibinfo {author}
  {\bibfnamefont{B.}~\bibnamefont{Velick\'y}},\ }%
  \bibfield{journal}{%
  \bibinfo {journal} {Phys. Rev. B}\ }%
  \textbf{\bibinfo {volume} {34}},\ \bibinfo {pages} {6933} (\bibinfo {year}
  {1986})%
  \bibAnnoteFile{NoStop}{Lipavsky}%
\bibitem{BaymConserving}%
  \BibitemOpen
  \bibfield{author}{%
  \bibinfo {author} {\bibfnamefont{G.}~\bibnamefont{Baym}},\ }%
  \bibfield{journal}{%
  \bibinfo {journal} {Phys. Rev.}\ }%
  \textbf{\bibinfo {volume} {127}},\ \bibinfo {pages} {1391} (\bibinfo {year}
  {1962})%
  \bibAnnoteFile{NoStop}{BaymConserving}%
\bibitem{LouiedWdU}%
  \BibitemOpen
  \bibfield{author}{%
  \bibinfo {author} {\bibfnamefont{S.}~\bibnamefont{Ismail-Beigi}}\ and\
  \bibinfo {author} {\bibfnamefont{S.~G.}\ \bibnamefont{Louie}},\ }%
  \bibfield{journal}{%
  \bibinfo {journal} {Phys. Rev. Lett.}\ }%
  \textbf{\bibinfo {volume} {90}},\ \bibinfo {pages} {076401} (\bibinfo {year}
  {2003})%
  \bibAnnoteFile{NoStop}{LouiedWdU}%
\bibitem{Kremp:Book}%
  \BibitemOpen
  \bibfield{author}{%
  \bibinfo {author} {\bibfnamefont{D.}~\bibnamefont{Kremp}}, \bibinfo {author}
  {\bibfnamefont{M.}~\bibnamefont{Schlanges}},\ and\ \bibinfo {author}
  {\bibfnamefont{W.-D.}\ \bibnamefont{Kraeft}},\ }%
  \emph{\bibinfo {title} {Quantum Statistics of Nonideal Plasmas}}\ (\bibinfo
  {publisher} {Springer},\ \bibinfo {address} {Berlin Heidelberg New York},\
  \bibinfo {year} {2005})%
  \bibAnnoteFile{NoStop}{Kremp:Book}%
\end{thebibliography}

%

\end{document}